\documentclass[12pt]{article}
\usepackage{multirow}
\usepackage{amsmath,amsfonts,amssymb,bm,slashed,bbm,mathrsfs}
\usepackage{graphicx,color}
\usepackage{graphicx}
\graphicspath{{./Pic/}}
\renewcommand{\Ref}[1]{(\ref{#1})}
\newcommand{\eq}[2]{\begin{align}\label{#1}#2\end{align}}

\newcommand{\nn}{\nonumber}

\newcommand{\pa}{\partial}

\newcommand{\sig}{\sigma}

\renewcommand{\Ref}[1]{(\ref{#1})}
\newcommand{\beao}{\begin{eqnarray*}}
\newcommand{\eeao}{\end{eqnarray*}}
\newcommand{\be}{\begin{equation}}
\newcommand{\ee}{\end{equation}}
\newcommand{\bea}{\begin{eqnarray}}
\newcommand{\eea}{\end{eqnarray}}
\newcommand{\beq}{\begin{eqnarray}}
\newcommand{\eeq}{\end{eqnarray}}%vesion with corrections 070221
\begin{document}
%\begin{center}
\title{Effective vertexes in magnetized quark-gluon plasma }
\author{
  V. Skalozub\thanks{e-mail: Skalozub@dnu.edu.ua} \\
{\small Oles Honchar Dnipro National University, 49010 Dnipro, Ukraine}}
\date{ }
%I. Gamolsky \\
%{\small Oles Honchar Dnipro National University, 49010 Dnipro, Ukraine}
%\end{center}
\date{\small}
%\title{ Quark propagation at  Polyakov's loop  background}
%\author{V. Skalozub,  A. Turinov}
%{\small Oles Honchar Dnipro National University, 49010 Dnipro, Ukraine}}
%\affil{Dnipro National University, Dnipro, Ukraine}
%\setcounter{Maxaffil}{Dnipro National University, Dnipro, Ukraine}
%\renewcommand\Affilfont{\itshape\small}
\maketitle
%%%%%%%%%%%%%%%%%%%%%%%%%%%%%%%%%%%%%%%%%%%%%%%%%%%%%%%%%%%%%%%%%
%\begin{center}
%{\bfseries
\begin{abstract}
In quark-gluon plasma (QGP), at  high temperatures $T$ the spontaneous generation of color magnetic fields, $b^3(T), b^8(T)  \not = 0$ (3, 8 are color indexes), and usual magnetic field $b(T)  \not = 0$ happens. Also,   the  Polyakov loop  and related to it the $A_0(T)$ condensate, which is solution to Yang-Mills imaginary time equations,     create.

Recently, with  the new type two-loop effective potential, which generalizes  the known integral representation for the  Bernoulli  polynomials and  takes into consideration the magnetic background, these effects were  derived.
   The corresponding      effective potential $W(T, b^3, b^8, b, A_0 )$  was  calculated  either in SU(2) gluodynamics or full quantum chromodynamics (QCD).  The values of magnetic field strengths at different temperatures were calculated  and the mechanism for stabilizing  the background  due to $A_0(T)$  was also  discovered.

    In  present paper, we  concentrate on the one-loop quark contributions.  In particular,  we derive the effective vertexes, which  couple  magnetic fields and $A_0$.  The vertexes result in new specific effects   signalling the creation  of  QGP  in  heavy ion collision experiments.
	
\end{abstract}
%\end{center}
\section{Introduction}
Deconfinement phase transition (DPT),  as well as the properties of  QGP, are widely investigated  for many years. Most results have been obtained in the lattice simulations because of the large coupling value $g \ge 1$ at the lower the phase transition temperature $T_d$. But at high temperatures due to asymptotic freedom the analytic methods are also reliable. They give a possibility for investigating  various phenomena in the plasma. Among them is the creation of gauge field condensates described by the classical solutions to field equations without sources. Only such type  fields could appear spontaneously inside the QGP. The well known ones are the so-called $A_0$ condensate, which is algebraically related to the Polyakov loop (PL) and  the chromomagnetic fields $b^3 = g H^3, b^8 = g H^8$ (3, 8 are color indexes of SU(3) group)  which are the Savvidy vacuum states at high temperature. These condensates result in numerous proper new effects  which could be the signals of the QGP. The  condensation of   $A_0$ alone is investigated by  different methods. For  recent works see, for instance, \cite{gaof21-103-094013} and references  therein.

All the mentioned condensates are the consequences of asymptotic freedom and follow from the important property  that asymptotic freedom at high temperature inevitably  results in an infrared instability at low one. The field condensation prevents such type instability that results in the formation of the physical  vacuum state.  In quantum field theory (QFT), the magnetic and the $A_0$ condensates are generated at different orders  in coupling constant (or the  number of loops) for the effective potential (EP) $W(T, b, b^3, b^8, A_0)$. So that they have different temperature dependencies and  play different  roles in the QGP dynamics.   For example, $A_0$ is generated at $g^4$ order in coupling constant and determined by the ratio of two- and one-loop  contributions to $W(A_0)$.  The fields $b(T), b^3(T), b^8(T)$ are generated in tree - plus one-loop - plus daisy approximation and also have the order $g^2$ in coupling constant. On the other hand, the contribution of $A_0$ at tree level equals  zero because it is a constant electrostatic potential.

The fields investigated below are an important topic towards a theory of confinement. The $A_0$-background is relevant because at finite temperature such field cannot be gauged away and is intensively investigated beginning with \cite{weis82-25-2667}. In the early 90-ies, two-loop contributions were calculated in QCD and with these,  the EP has  non-trivial minimums and related condensate fields (see, for instance, \cite{skal94-57-324}, \cite{skal21-18-738}). They form a hexagonal structure in the plane of the  color components $A_0^3$ and $A_0^8$ of the background field.

 A common generation of both fields was studied analytically in \cite{bord22-82-390}. Here, new  representation  generalizing the known integral representation for the Bernoulli polynomials, was worked out, which  admits introducing either $A_0$ or any $b$ fields up to two-loop order.
 The magnetic fields considerably change the spectra of quarks and gluons as well.
 So, new phenomena have to be realized.  The PL as well as  $A_0(T)$ are the order parameters for the deconfinement phase transition.    At low temperature they equal zero. At high temperature they become nonzero. The same concerns the spontaneously created magnetic fields.

The SU(3) gauge group can be presented as three  SU(2) groups. So, the most results  are relevant.
In   SU(2), the EP in the background $R_\xi$ gauge reads \cite{bord22-82-390}:
\eq{W2}{ W^{SU(2)}_{gl} &=B_4(0,0)+2B_4\left(a,b\right)
\\\nn&~~~	+2{g^2}\left[
	B_2\left(a,b\right)^2
	+2 B_2\left(0,b\right) B_2\left(a,b\right)	\right]
	-4{g^2}(1-\xi) B_3\left(a,b\right)B_1\left(a,b\right)
}
with the notation
\eq{ab}{a=\frac{x}{2}=\frac{g A_0}{2\pi T},~~~b=gH_3^3.
}
The chromomagnetic field is directed along third directions in coordinate and color spaces.
Since we work at finite temperature, $ W_{gl}$ is equivalent to the free energy.

The functions $B_n(a,b)$ are defined by
\eq{3}{ B_4(a,b) &= T\sum_\ell\int\frac{dk_3}{2\pi}\frac{b}{4\pi^2}\sum_{n,\sig}
	\ln\left(\left(2\pi T(\ell+a)\right)^2+k_3^2+b(2n+1+\sig-i0)\right),
\\\nn
B_3(a,b) &=
T\sum_\ell\int\frac{dk_3}{2\pi}\frac{b}{4\pi^2}\sum_{n,\sig}
\frac{\ell+a}{\left(2\pi T(\ell+a)\right)^2+k_3^2+b(2n+1+\sig-i0)}
\\\nn
	 B_2(a,b) &= T\sum_\ell\int\frac{dk_3}{2\pi}\frac{b}{4\pi^2}\sum_{n,\sig}
	\frac{1}{\left(2\pi T(\ell+a)\right)^2+k_3^2+b(2n+1+\sig-i0)},
\\\nn		
	B_1(a,b) &=
	T\sum_\ell\int\frac{dk_3}{2\pi}\frac{b}{4\pi^2}\sum_{n,\sig}
	\frac{\ell+a}{\left(\left(2\pi T(\ell+a)\right)^2+k_3^2+b(2n+1+\sig-i0)\right)^2}.
}
In eq.\Ref{W2}, $\xi$ is gauge fixing parameter,  the summations run  $n=0,1,\dots$, $\sig=\pm2$ and $\ell$ runs over all integers. The $'-i0'$-prescription  defines the sign of the imaginary part for the tachyon  mode.
These formulas and eq.\Ref{W2} are the generalization of  corresponding two-loop expressions in \cite{enqv90-47-291},  \cite{bely91-254-153},   \cite{skal92-7-2895}  and also \cite{skal21-18-738}
for including a magnetic field.  We note  the relations
\eq{3a}{B_3(a,b) &= \frac{1}{4\pi T}\pa_a B_4(a,b),&
		B_1(a,b) &= \frac{-1}{4\pi T}\pa_a B_2(a,b)
}
proper to the Bernoulli polynomials.

For $b = 0$ we have to replace  $\frac{b}{4\pi^2}\sum_{n,\sig}\to\int\frac{d^2k}{(2\pi)^2}$ and get
\eq{1.1}{ 	&&B_4(a,0)=\frac{2\pi^2 T^4}{3} B_4(a),
	 	~~	B_3(a,0)=\frac{2\pi T^3}{3}	B_3(a), \\ \nonumber
  		 &&B_2(a,0)=\frac{T^2}{2}	B_2(a),
	~~~~~~~	B_1(a,0)=-\frac{ T}{4\pi B_1(a)},
}
where $B_n(a)$ are the Bernoulli polynomials, periodically continued.
The special values for $a=0$ are
\eq{4}{ B_4(0,0)  = -\frac{\pi^2T^4}{45},
	~~~   B_3(0,0)= 0,
	~~~   B_2(0,0)= \frac{T^2}{12},
	~~~   B_1(0,0)= \frac{T}{8\pi}.
}
These formulas hold for $T>0$.

 Note also that the function $B_4(a,b)$ describes the one loop contribution  and the others give the two loop part of the EP.
The motivation for the above choice of notations  is that  the functions $B_n(a,b)$, \Ref{3} are  the corresponding mode sums without additional factors. More details about this representation as well as the renormalization and the case of $T = 0$ are given in \cite{bord22-82-390}. Above expressions with whole $l = \pm 1, \pm 2,...$ correspond to boson contributions. For fermions $ l = \pm (1 + 1/2), \pm (2 + 1/2),...$.
\section{One loop approximation for EP}
As mentioned in the previous section, the spontaneous creation of $A_0$ field in the Matsubara imaginary time formalism
 happens in two-loop approximation presented in eq.(1).  In one-loop order, only magnetic fields are generated. The latter part at finite temperature is described by the expression $B_4 (a,b)$ which will be the main object below. In  case of a number of   fields it produces numerous specific interactions between them (due to vacuum fluctuations).

To explain the origin of this object we start with   $b = 0$ case (see, for example, \cite{skal94-57-324} which is used in what follows). In Appendixes $A_1, A_2$ of it the Feynman  rules and the integral representations for the Bernoulli polynomials are placed. For us, the  QCD condensates generated by fermion  loops  are needed: $c_1 = g ( A^3_0 + A^8_0/\sqrt{3})/2,$ $c_2 = g ( - A^3_0 + A^8_0/\sqrt{3})/2, c_3 = g ( - A^8_0/\sqrt{3})$. Here, 3 and 8 are  internal SU(3) group color indexes.

The basic integral representation for $B_4(\frac{C\beta}{2 \pi})$, $\beta = 1/T$ reads (we changed notation $k_i \to p_i $ to be in correspondence with \cite{skal94-57-324}),

\eq{B_4}{ \int d p \ln p^2_c = \frac{2 \pi^2}{3 \beta^4}
  B_4(\frac{C \beta}{2 \pi} ).}
Here, $\int d p  = \frac{1}{\beta} \sum_{p_0} \int \frac{d^3 p}{ (2 \pi)^3},$ and $ p^2_c = (p_0 + c)^2 + \vec{p}^2 .$ Summation over $p_0 $ runs from minus to plus infinity with the values $ p_0 = \frac{2 \pi l}{\beta}$ for bosons, $ p_0 = \frac{2 \pi (l + 1/2)}{\beta}$ for fermions and the value of c = $c_i$.

To obtain formulas of eq.(3) we have to make the substitutions in the integral representation for $B_4(c_i, \vec{p})$: $ \int d p \to \int \frac{d p_3}{(2 \pi)} \frac{g H}{(2 \pi)^2} \sum_{(n, \sigma)} $, where $ n = 0, 1,2,... $ is  a Landau level  number and $\sigma = \pm 1 $ is spin number. This is in accordance with energy spectrum of  charged spin $1/2 $ particle in constant magnetic field: $\epsilon_p^2 = p_3^2 + g H ((2 n + 1) -\sigma ) + m^2$. $p_3$ is  momentum along field direction $H_3 = const$. Here the kind of magnetic field is insufficient. We assume  that all the generated magnetic fields are directed along third axis in coordinate space. As it is occurred, for parallel fields the minimum of the effective potential is lower.    In the ground state $n = 0, \sigma = 1$, and the particle energy is $\epsilon_p^2 = p_3^2 + m^2$. This is so called Low landau level (LLL) approximation. In strong fields it significantly simplifies calculations and gives  very good results. Therefore, it will be used in what follows.

Our next steps are the following. First, we  calculate the derivative with respect to $\epsilon_p^2 $ of the $B_4(c_i, \epsilon_p^2)$. Second, we sum up the series over $ p_0 $. Then we integrate over $\epsilon_p^2$ and obtain the potential of the effective produced interactions between magnetic and $A_0$ fields. This nontrivial procedure results in the new type effective vertexes for high temperature QGP. Finally, we summarize our results and discuss prospects for future researches.
\section{Calculation  series over  $p_0$}
Now, we calculate the temperature sum over $p_0$ proceeding in  two steps. First we calculate derivative of the $B_4(c_i, \epsilon_p^2)$ eq.(3) with respect to $\epsilon_p^2$ which in a magnetic field background is realized by  substituting  $ \vec{p}^2 \to \epsilon_p^2$. Remind that $ \epsilon_p^2 = p_3^2 + g H ((2n + 1) - \sigma)$ and in LLL approximation $ n = 0, \sigma = 1, \epsilon_p^2 = p_3^2 + m^2_q = \epsilon^2_{pL} , m_q$ is quark mass. Below we write the parameter $\epsilon_p^2$ for all cases, where it is clear.

We write for the left hand side of $B_4(c_i, \epsilon_p^2) = I_0 $ and obtain
\eq{I1}{I_1 = \frac{ d I_0}{d \epsilon_p^2} = \frac{1}{\beta} \sum_{p_0}\int \frac{d p_3}{(2 \pi)} \frac{g H}{(2 \pi)^2} \sum_{(n, \sigma)} \frac{1}{(p_0 + C)^2 + \epsilon_p^2}.}
Here, C stands for one of $c_i$ written above and $p_0$ corresponds to fermions.  To calculate series over $p_0$ we use the standard integral representation for fermions
%\frac{1}{\beta} \sum_{p_0}
\eq{Spo}{ \frac{1}{\beta} \sum_{p_0} f(p_0) = -  \frac{1}{2} \sum_{j} Res[f(\omega)\tan (\frac{\beta \omega}{2})],}
where summation is over poles $\omega_j$.

In our case, the poles are, $\omega_1 = - C + i\epsilon_p,~~ \omega_2 = - C -i \epsilon_p$. By using these values we obtain the result
\eq{I2}{I_2 = \frac{1}{2 \epsilon_p} (\frac{\sinh(\epsilon_p \beta)}{ \cos(C \beta) + \cosh(\epsilon_p \beta)} - 1) .}
In this expression we have subtracted $1$ to separate a zero temperature contribution. In fact, in this expression $\epsilon_p = (p_3^2 + m^2)^{1/2}$ because we turn to the LLL approximation and not calculated complete series over $n, \sigma.$
\section{Conclusions}
As  final step, we integrate over $\epsilon_p^2$ as a parameter. The result is 
\eq{I0f}{I_{2F} = - \epsilon_p + \frac{1}{\beta} \ln[\cosh(\epsilon_p \beta)+ \cos(C \beta)] .}
This expression has to be inserted in the integral over $d p_3$ eq. (8). We obtain,
 \eq{I0fr}{ V_{H,A_0} = \frac{g H}{(2 \pi)^2} \int
\frac{d p_3}{(2 \pi)} (- \epsilon_p + \frac{1}{\beta} \ln[\cosh(\epsilon_p \beta)+ \cos(C \beta)] ),}
where $C = c_i$ written in the second section. It describes effective interactions of different kind magnetic fields $ H^3, H^8$ and usual magnetic field H  with $A_0$ condensates in high temperature $QGP$.

 Derived potential generates new type effective vertexes and produces new effects in heavy ion collisions. Investigation of them will be reported in other place.

\end{document}